\documentclass[8.5pt,twoside,twocolumn]{article}
\usepackage[super,sort&compress,comma]{natbib} 
\usepackage{mhchem}
\usepackage{times,mathptmx}
\usepackage{sectsty}
\usepackage{balance} 

\usepackage{graphicx} 
\usepackage{lastpage}
\usepackage[format=plain,justification=raggedright,singlelinecheck=false,font=small,labelfont=bf,labelsep=space]{caption} 
\usepackage{fancyhdr}
\begin{document}

\thispagestyle{plain}
\fancypagestyle{plain}{
\fancyhead[L]{\includegraphics[height=8pt]{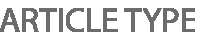}}
\fancyhead[C]{\hspace{-1cm}\includegraphics[height=20pt]{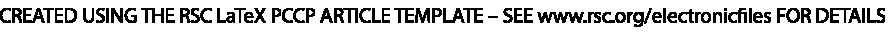}}
\fancyhead[R]{\includegraphics[height=10pt]{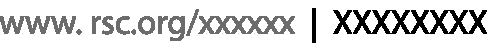}\vspace{-0.2cm}}
\renewcommand{\headrulewidth}{1pt}}
\renewcommand{\thefootnote}{\fnsymbol{footnote}}
\renewcommand\footnoterule{\vspace*{1pt}%
\hrule width 3.4in height 0.4pt \vspace*{5pt}} 
\setcounter{secnumdepth}{5}

\makeatletter 
\def\subsubsection{\@startsection{subsubsection}{3}{10pt}{-1.25ex plus -1ex minus -.1ex}{0ex plus 0ex}{\normalsize\bf}} 
\def\paragraph{\@startsection{paragraph}{4}{10pt}{-1.25ex plus -1ex minus -.1ex}{0ex plus 0ex}{\normalsize\textit}} 
\renewcommand\@biblabel[1]{#1}            
\renewcommand\@makefntext[1]%
{\noindent\makebox[0pt][r]{\@thefnmark\,}#1}
\makeatother 
\renewcommand{\figurename}{\small{Fig.}~}
\sectionfont{\large}
\subsectionfont{\normalsize} 

\fancyfoot{}
\fancyfoot[LO,RE]{\vspace{-7pt}\includegraphics[height=9pt]{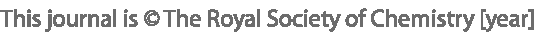}}
\fancyfoot[CO]{\vspace{-7.2pt}\hspace{12.2cm}\includegraphics{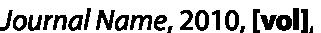}}
\fancyfoot[CE]{\vspace{-7.5pt}\hspace{-13.5cm}\includegraphics{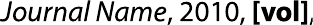}}
\fancyfoot[RO]{\footnotesize{\sffamily{1--\pageref{LastPage} ~\textbar  \hspace{2pt}\thepage}}}
\fancyfoot[LE]{\footnotesize{\sffamily{\thepage~\textbar\hspace{3.45cm} 1--\pageref{LastPage}}}}
\fancyhead{}
\renewcommand{\headrulewidth}{1pt} 
\renewcommand{\footrulewidth}{1pt}
\setlength{\arrayrulewidth}{1pt}
\setlength{\columnsep}{6.5mm}
\setlength\bibsep{1pt}

\twocolumn[
  \begin{@twocolumnfalse}
    \noindent\LARGE{\textbf{Collective Ratchet Effects and Reversals for Active
        Matter Particles on Quasi-One-Dimensional Asymmetric Substrates }}
\vspace{0.6cm}

\noindent\large{\textbf{Danielle McDermott,\textit{$^{1,2}$}  
Cynthia J. Olson Reichhardt,$^{\ast}$\textit{$^{1}$} 
and Charles Reichhardt\textit{$^{1}$}}}\vspace{0.5cm}

\noindent\textit{\small{\textbf{Received Xth XXXXXXXXXX 20XX, Accepted Xth XXXXXXXXX 20XX\newline
First published on the web Xth XXXXXXXXXX 200X}}}

\noindent \textbf{\small{DOI: 10.1039/b000000x}}
\vspace{0.6cm}

\noindent \normalsize{
Using computer simulations,  we study a two-dimensional system of
sterically interacting self-mobile run-and-tumble disk-shaped particles
with an underlying
periodic quasi-one-dimensional asymmetric substrate, and show that
a rich variety of
collective active ratchet behaviors arise
as a function of particle density, activity, substrate strength, and
substrate period.
The ratchet efficiency is nonmonotonic since
the ratcheting is enhanced by increased activity but diminished
by the onset of self-clustering of the active particles.
Increasing the particle density decreases the ratchet efficiency for weak
substrates but increases the ratchet efficiency for strong substrates
due to collective hopping events.
At the highest particle densities, the ratchet motion is
destroyed by a self-jamming effect.
We show that it is possible to realize reversals of the ratchet effect, 
where the net flux of particles is along the hard rather than the easy
direction of the substrate asymmetry.
The reversals occur in the strong substrate limit when multiple rows of active particles 
can be confined in each substrate minimum, permitting emergent particle-like
excitations to appear that experience an inverted effective substrate potential.
We map out a phase diagram of the forward and reverse ratchet effects as a
function of the particle density, activity, and substrate properties.
}
\vspace{0.5cm}
  \end{@twocolumnfalse}
]

\section{Introduction}

\footnotetext{\textit{
$^1$~Theoretical Division,
    Los Alamos National Laboratory, Los Alamos, New Mexico 87545 USA.
    Fax: 1 505 606 0917; Tel: 1 505 665 1134; E-mail: cjrx@lanl.gov}}
\footnotetext{\textit{
$^2$~Department of Physics, Wabash College, Crawfordsville, Indiana 47933 USA.}} 

A nonequilibrium assembly of particles subjected to a driving force that,
by itself, produces no
net motion can exhibit a net dc drift called a ratchet effect
when enough symmetries are broken, such as by the introduction of
an asymmetric substrate.
A wide variety of ratchet effects are possible, including a rocking
ratchet for particles driven over an asymmetric substrate by an ac force,
or a flashing ratchet for thermally fluctuating particles interacting with a
substrate that is switched on and off periodically \cite{1,2,3}.
Ratchet effects produced using asymmetric substrates have been studied
in a wide range of systems including
colloidal particles \cite{4},
vortices in type-II superconductors \cite{5,6},
magnetic domain walls \cite{7,8},
polymers \cite{9,10}, 
granular matter \cite{11,12}, and cold atoms \cite{13}.
Systems with symmetric substrates can also exhibit ratchet effects
if additional asymmetry is introduced, such as from an
asymmetric external driving force
\cite{14,15,16}.
For overdamped non-interacting particles 
on an asymmetric substrate, 
the normal ratchet effect creates a particle drift  
in the easy flow direction of the substrate asymmetry.
Interacting particle systems can exhibit a reversal,  
or even multiple reversals, of the ratchet effect in which the net drift
is in the hard direction
of the substrate asymmetry \cite{17}.
Ratchet reversals
have been observed in rocking ratchets
for interacting superconducting vortices in asymmetric pinning arrays
as a function of vortex density or ac drive amplitude
\cite{18,19,20,21,22}.
While in these systems some form of external driving must be applied to
produce the ratchet effect,
for self-propelled particles, known as active matter \cite{23,24,New1}, 
ratchet effects can appear in the absence
of external driving \cite{25,26,27,N1,New30}.
  
Rectification effects in active matter systems were first
observed for 
run-and-tumble swimming bacteria moving through an array of
funnel-shaped barriers \cite{25}.
The initially uniformly distributed bacteria
become concentrated on the easy flow side of the funnels over time, 
while non-swimming bacteria undergo no rectification \cite{25}.
Subsequent simulations of run-and-tumble particles in similar funnel geometries
produced a similar ratchet effect caused by the interaction of the running particles
with the asymmetric funnel walls, while the ratchet effect disappeared in the
Brownian limit of very short running times \cite{26}.
Further studies showed that this rectification effect
depends on the nature of the particle interactions 
with the barriers; 
a ratchet effect can occur 
when detailed balance is broken, 
but is absent 
when the particles scatter elastically from the barriers \cite{27}.
Active ratchets have been studied for 
a variety of self-driven systems in the presence of 
asymmetric substrates \cite{N,28,29,30} or 
asymmetric obstacles \cite{31,32,33}, as well as for 
more complicated self-driven systems such as active ellipsoids \cite{34}, 
active polymers \cite{35}, and self-driven Janus particles \cite{36}
on asymmetric substrates.
Other studies have shown how active ratchet effects
can be exploited to transport non-active colloidal cargo \cite{37}, 
rotate asymmetric gears immersed in active matter \cite{38,New2}, 
capture active particles with asymmetric traps \cite{L1},
and direct the motion of asymmetric obstacles 
in active matter baths \cite{39,40}.

\begin{figure}
  \centering
\includegraphics[width=0.4\textwidth]{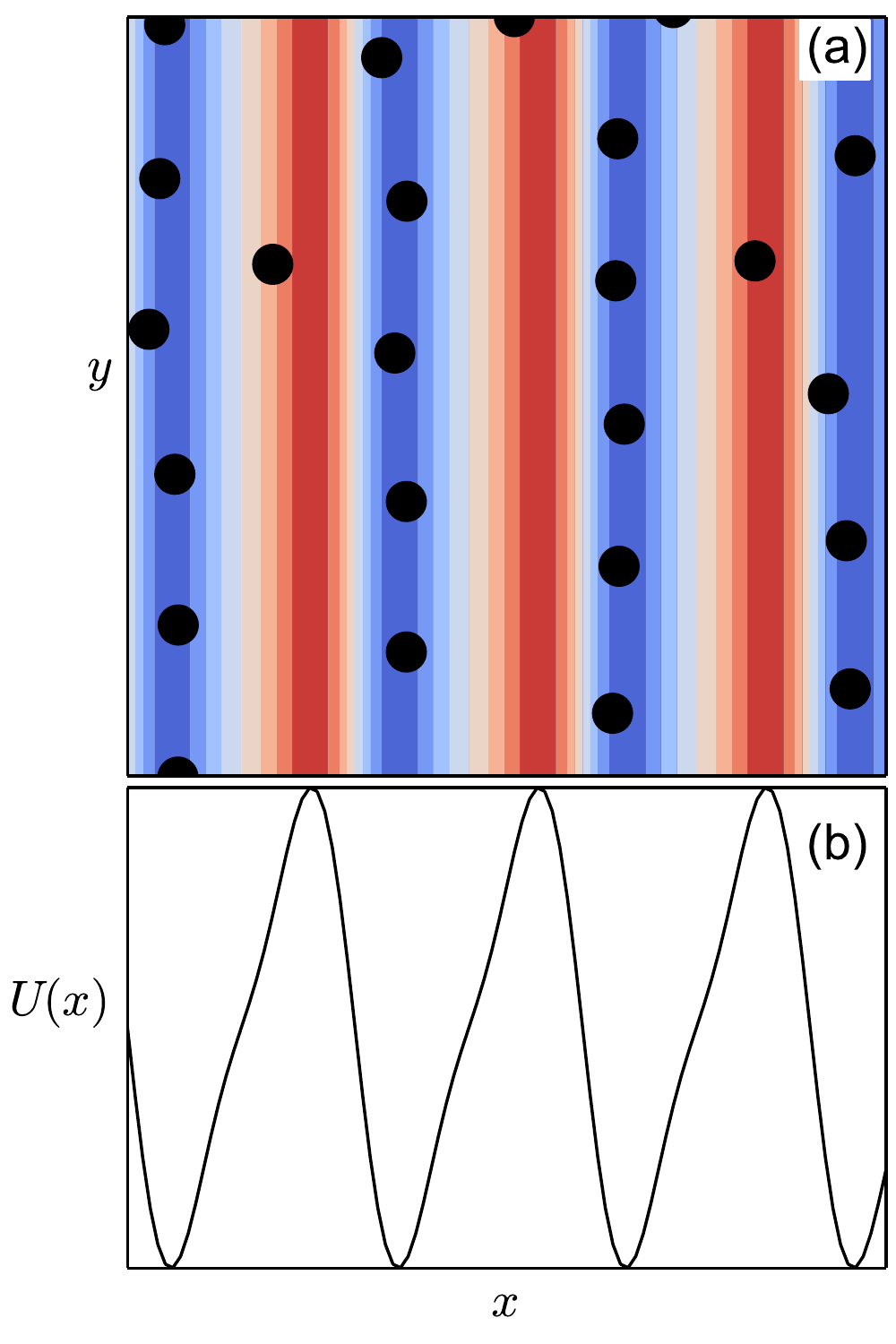}
\caption{ (a) Top view of a portion of the asymmetric potential $U(x)$, where
  red (blue) shading indicates high (low) potential energy.
  The black circles schematically show the locations of the self-propelled particles.
  (b) The corresponding shape of the potential $U(x)$ vs $x$.  
}
\label{fig:1}
\end{figure}

When interactions between
active matter particles are included, 
dynamical effects
such as self-trapping \cite{41} and 
clustering effects \cite{42,43,44,45}
arise 
even in the absence of a substrate.
Recent simulations of disk-shaped active particles driven through
an array of obstacles 
show that the drift velocity of the particles
initially increases with increasing activity,
but decreases with increasing activity
once self-clustering or self-jamming begins to occur \cite{46}.
It might be expected that
active matter ratchet effects would generally diminish when
particle-particle interactions are introduced,
as observed by
Wan \textit{et al.} 
for run-and-tumble particles moving through a funnel array, 
where the ratchet effect
decreased with increasing particle density \cite{26}.
This is, however, not always the case.
For
active matter particles
on asymmetric substrates,
reversals in the ratchet effect have been observed for
interacting particles obeying a flocking or Vicsek model \cite{47}
that move through an array of funnel barriers \cite{48} 
as well as in experiments on eukaryotic cells crawling through
asymmetric channels \cite{49}. 
For simpler systems such as self-propelled disks 
or rods on asymmetric substrates, 
ratchet reversals have not yet been observed.

In this work we examine a two-dimensional (2D) system 
of sterically interacting run-and-tumble disk-shaped particles moving over an
asymmetric
quasi-one-dimensional periodic substrate.
At low particle densities,
a ratchet effect occurs 
in the easy flow direction of the substrate asymmetry.
For weak substrates, 
the ratchet efficiency increases with 
increasing run time, and it
decreases with increasing particle density when self-clustering occurs.
For stronger substrates, 
the ratchet efficiency is a nonmonotonic function of
particle density and run time.
When the substrate is strong enough to confine multiple rows of particles
in each potential minimum, multibody collisions can occur
that push particles over the substrate barriers,
generating a reversed ratchet effect with motion in the hard direction
of the substrate asymmetry.  This reversed ratchet effect is
suppressed
at high particle densities when self-clustering occurs, and it also disappears
for very strong substrates when the particles form one-dimensional rows in which
multibody collisions do not occur.
A transition from a reverse to a normal ratchet effect can occur
when the run time is increased.
We describe the direction and efficiency 
of the ratchet effect in a series of phase diagrams 
as functions of the particle density, run time, substrate strength, 
and substrate periodicity.

\begin{figure*}
\includegraphics[width=\textwidth]{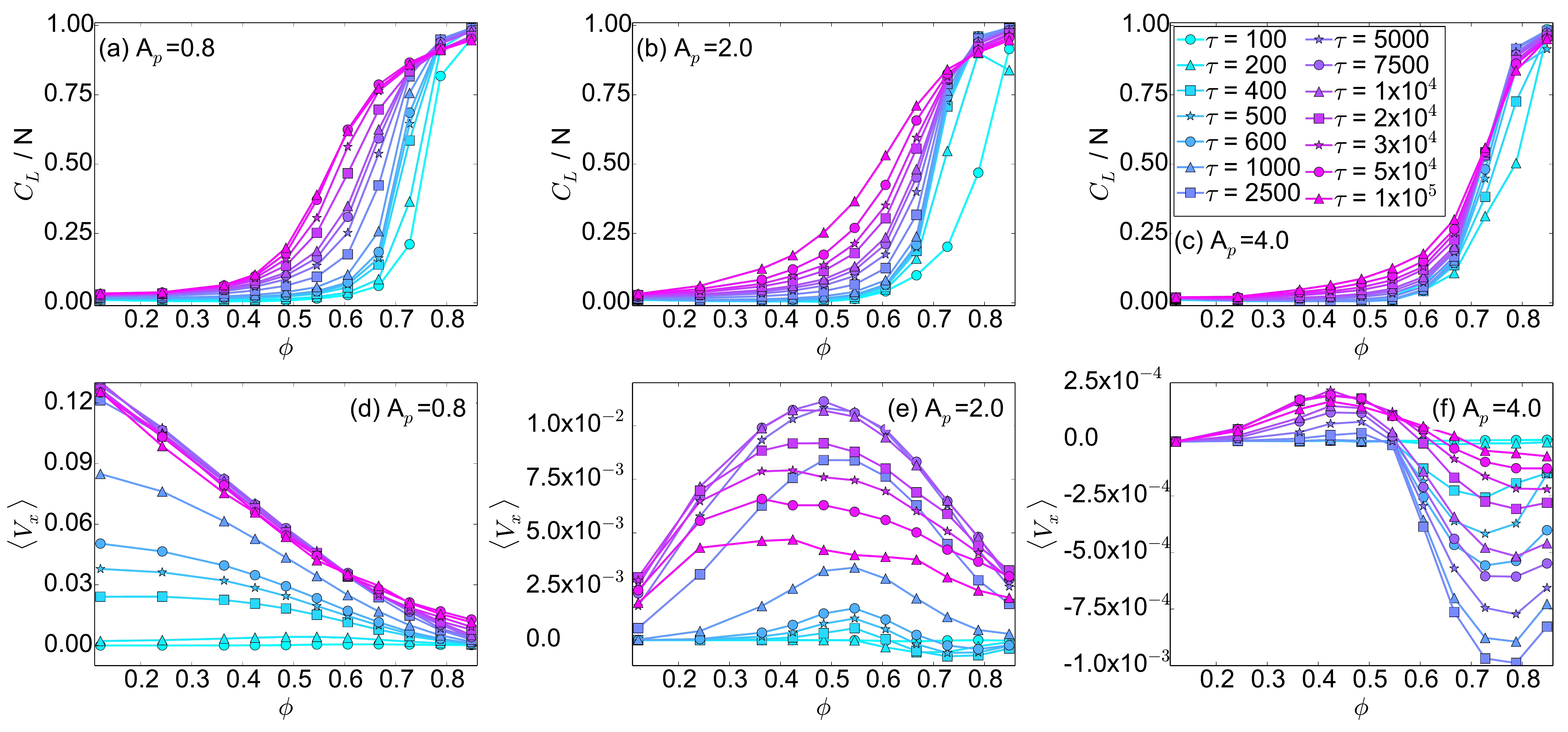}
\caption{(a,b,c) The average cluster size $C_{L}/N$ vs particle density $\phi$. 
(d,e,f) The average particle velocity $\langle V_{x} \rangle$ vs $\phi$. 
  The color code, shown in panel (f), indicates different
  run times $\tau=100$, 200, 400, 500, 600, 1000, 2500, 5000, 7500,
    $1\times 10^4$, $2\times 10^4$,  $3\times 10^4$, $5\times 10^4$, and $1\times 10^5$.
  (a,d) At $A_{p} = 0.8$,
there is a normal ratchet effect with an efficiency that
decreases with increasing $\phi$.
(b,e) At $A_{p} = 2.0$,
the efficiency of the normal ratchet effect is nonmonotonic,
so that the ratcheting is optimized for a midrange value of $\phi$.
(c,f) At $A_{p} = 4.0$,
there is a crossover from a normal to a reverse
ratchet effect with increasing $\phi$.
}
\label{fig:2}
\end{figure*}

\section{Simulation and System}
We consider a 2D system of size $L \times L$
with periodic boundary conditions in the $x$- and $y$-directions containing
$N$ self-propelled disk-shaped particles.
The steric particle-particle interactions are modeled 
as a harmonic repulsive potential which drops to zero beyond the
particle radius $r_{d}$.
We take $r_{d} = 0.5$ and $L = 36$ in dimensionless simulation length units. 
The particle density $\phi$ is given by the total
fraction of the sample area covered
by the particles,
$\phi = N\pi r^2_{d}/L^2$. 
The highest possible particle density in 2D is
a triangular solid with $\phi=0.9$.
The dynamics of a particle $i$ is governed by the following
overdamped equation of motion:
\begin{equation}
\eta \frac{d {\bf r}_{i}}{dt} =
 {\bf F}_i^{m} + {\bf F}^{in}_i + {\bf F}^{sub}_{i} \ .
\end{equation}
Here $\eta$ is the damping constant, 
which is set equal to unity. 
The self-propulsion is modeled 
using
run-and-tumble dynamics in which the motor force ${\bf F}_i^{m}$
is fixed to a randomly chosen direction during a running time
$\tau$, after which a new randomly chosen direction is selected for the
next running time $\tau$.
We take the magnitude of the motor force to be $F_m=1.0$.
In the absence of any other interactions,
during a single run interval 
a particle travels a distance
$R_{l} = F_{m}\tau \Delta t$,
where $\Delta t=0.002$ is the size of the simulation time step.
The steric particle-particle
interaction force is
$F^{in}_{i} = \sum^{N}_{i\ne j}k(2r_{d} - |{\bf r}_{ij}|)
\Theta(2r_{d} - |{\bf r}_{ij}|){\hat {\bf r}}_{ij}$
where ${\bf r}_{ij}={\bf r}_i-{\bf r}_j$, ${\hat {\bf r}}_{ij}={\bf r}_{ij}/|{\bf r}_{ij}|$,
$\Theta$ is the Heaviside step function, and the spring constant $k=30$.
The substrate
force
${\bf F}_i^{sub} = -\nabla U(x_i){\hat {\bf x}}$ 
arises from an asymmetric potential of the form
\begin{equation}
U(x) = U_{0}[\sin(2\pi x/a) + 0.25\sin(4\pi x/a)]
\end{equation} 
where $a$ is the substrate period and the
substrate strength is defined to be $A_{p} = 2\pi U_{0}/a$.
Unless otherwise noted, we take $a=1.0$.
A small section of $U(x)$ is illustrated in
Fig.~\ref{fig:1}.
To quantify the ratchet effect, 
we measure the net velocity per particle 
in the $x$ direction, 
$\langle V_{x} \rangle  = N^{-1}\sum_{i = 1}^{N} {\bf v}_{i}\cdot \hat{\bf x},$
where ${\bf v}_i$ is the velocity of particle $i$.  We average
$\langle V_{x}\rangle$
over at least $10^7$ simulation time steps
to ensure that we are obtaining a steady state measurement. 
We vary 
$\phi$, $A_{p}$, $a$, and $\tau$ 
and measure the resulting
direction and efficiency of the ratcheting behavior.

\begin{figure}
\includegraphics[width=3.5in]{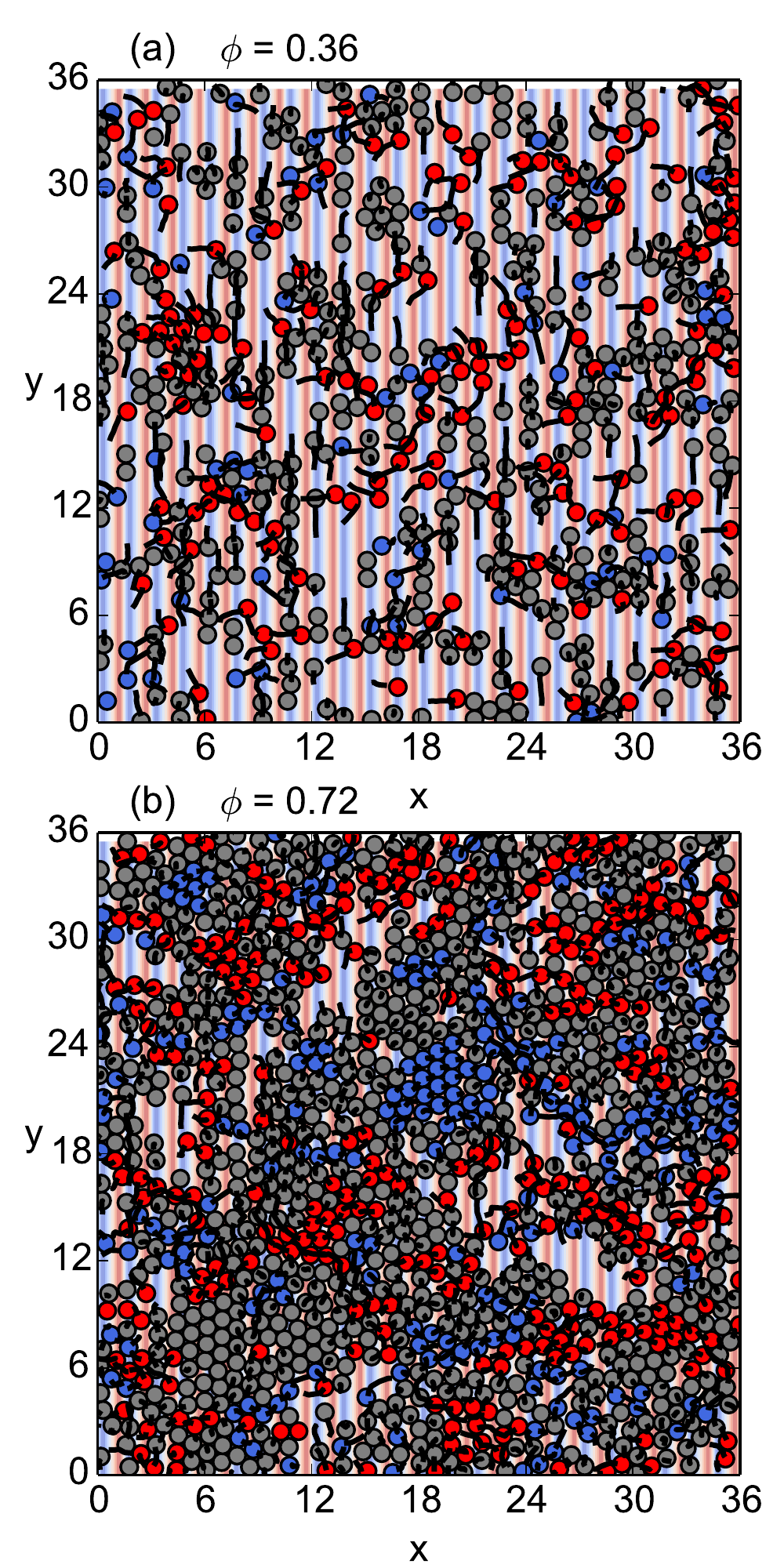}\\
\caption{ Active particle positions (circles) and
  trajectories (black lines), along with the underlying
  substrate potential (red and blue lines),
at $A_{p} = 0.8$ and $\tau=10^5$ for the
system in Fig.~\ref{fig:2}(a,d).
Red particles have moved a distance $\Delta x\geq a/4$
in the forward (positive $x$)
direction during the illustrated time interval, while blue particles have
moved a distance $\Delta x \geq a/4$
in the reverse (negative $x$) direction;
gray particles have moved a distance $\Delta x<a/4$.
(a) At $\phi = 0.36$, there 
is a large normal ratchet effect and few clusters are present.
(b)
At $\phi = 0.72$,
self-cluster formation suppresses the ratchet effect.
}
\label{fig:3}
\end{figure}

\section{Results}

In Fig.~\ref{fig:2}(a) we plot the
normalized size of the largest particle cluster
$C_{L}/N$ versus $\phi$ 
for varied run times $\tau$ at a fixed substrate strength of 
$A_{p} = 0.8$.
A group of $n$ particles that are all in physical contact with each other is
defined to be a cluster of size $C_L=n$.
We measure $C_L$ using
the
algorithm described in \cite{H}, and
define the system to be in a self-clustering state when
$C_{L}/N > 0.5$.
As $\tau$ increases, the onset of self-clustering shifts to
lower values of $\phi$, 
in agreement with earlier studies showing 
that clustering
occurs at lower densities 
when the run time \cite{46} or
the persistence length \cite{42,43} of the
motor force is increased.
Figure~\ref{fig:2}(d) shows 
the corresponding values of $\langle V_{x} \rangle$ versus $\phi$.
For the lowest value of $\tau = 100$,
the system is in the Brownian limit
and $\langle V_{x} \rangle = 0$ for all values of $\phi$, 
while at higher run times, 
the system exhibits a normal ratchet effect.  
In this weak substrate regime, we find that
$\langle V_{x} \rangle$ decreases 
with increasing $\phi$ and
increases with increasing $\tau$, with a saturation
for $\tau>5000$ as shown
in Fig.~\ref{fig:2}(d).
This result is in agreement with the studies of Wan \textit{et al.} 
on interacting active particles in funnel arrays,
where the magnitude of the ratchet effect increases
with increasing $\tau$ before reaching a plateau at large $\tau$,
and decreases with increasing particle density \cite{26}.
For the system in Fig.~\ref{fig:2}(a,d),
at low $\phi$ very few particle-particle collisions occur,
making the behavior similar to that of a single particle, which
moves further along the easy direction of the substrate
asymmetry than along the hard direction 
since $F^{m} > A_{p}$.
As $\phi$ increases, two effects combine to reduce the
magnitude of the forward ratchet effect.
The particles collide more often, producing a more thermal
distribution of particle motion, while individual particles are
more likely to become trapped inside a cluster, therefore
becoming unavailable for hopping over the substrate barrier.

\begin{figure}
\includegraphics[width=3.5in]{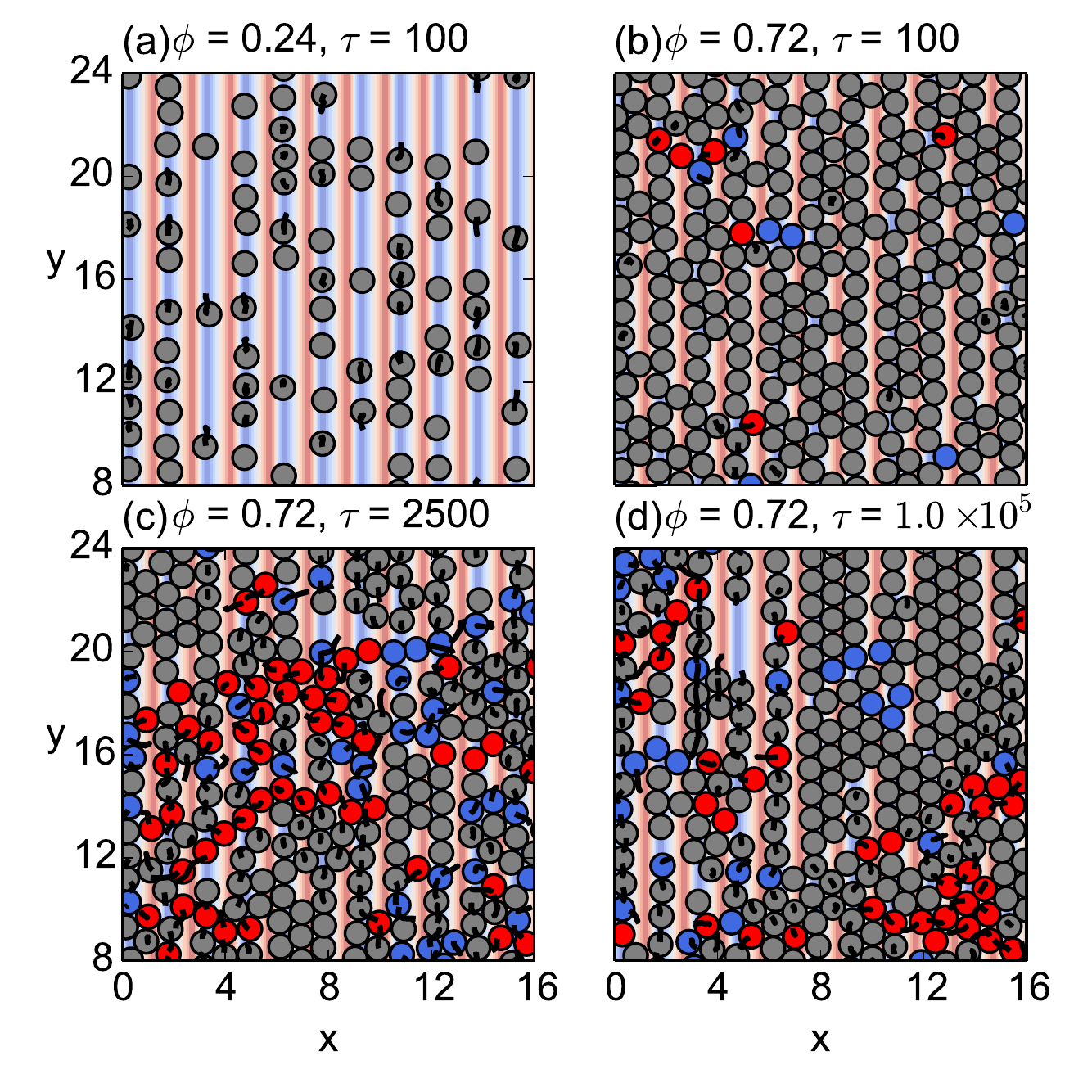}
\caption{Active particle positions (circles) and 
  trajectories (black lines),  along with the underlying
  substrate potential (red and blue lines), in a portion of the sample
  for the system in Fig.~\ref{fig:2}(b,e) with $A_{p} = 2.0$.
  Particles are colored as in Fig.~\ref{fig:3}.
  (a) At $\phi = 0.24$ and $\tau = 100$, 
$\langle V_{x} \rangle = 0$ and the particles are
trapped in the substrate minima.
(b) At $\phi = 0.72$ and $\tau = 100$, a uniform pinned
state forms with $\langle V_{x} \rangle = 0.$
(c) At $\phi = 0.72$ and $\tau = 2500$, a normal
ratchet effect occurs.
(d) At $\phi = 0.72$  and $\tau = 1\times 10^5$, the occurrence
of clustering reduces the magnitude of the normal ratchet effect.
}
\label{fig:4}
\end{figure}

In Fig.~\ref{fig:3} we show the 
particle positions and trajectories for the weak
substrate system from Fig.~\ref{fig:2}(a,d) with $A_p=0.8$
at $\tau=10^5$.
At $\phi=0.36$, there is a large normal ratchet effect
and Fig.~\ref{fig:3}(a) shows
that few clusters are present.
Particles that have moved a distance $\Delta x \geq a/4$
in the easy (positive $x$) direction are colored
red, while those that have moved a distance $\Delta x \geq a/4$
in the hard (negative $x$)
direction are colored blue.
As indicated in Fig.~\ref{fig:3}(a), particles are frequently able to
hop over the substrate barriers, and
due to the substrate asymmetry, hops in the easy direction occur more
often than hops in the hard direction, leading to a net normal ratchet effect.
At $\phi=0.72$, shown in
Fig.~\ref{fig:3}(b),
strong self-clustering is present and
the normal ratchet effect is much weaker. 
Due to the clustering,
the particles tend to move collectively at this density,
and large regions of the system become jammed, with
particles unable to hop over substrate barriers in either direction.
The high particle density also tends to nullify the effectiveness of the
substrate asymmetry, so that the few particles that are still able to
hop over barriers have a nearly equal probability of hopping in the
easy direction as in the hard direction, destroying the ratchet effect.

\begin{figure}
\includegraphics[width=3.5in]{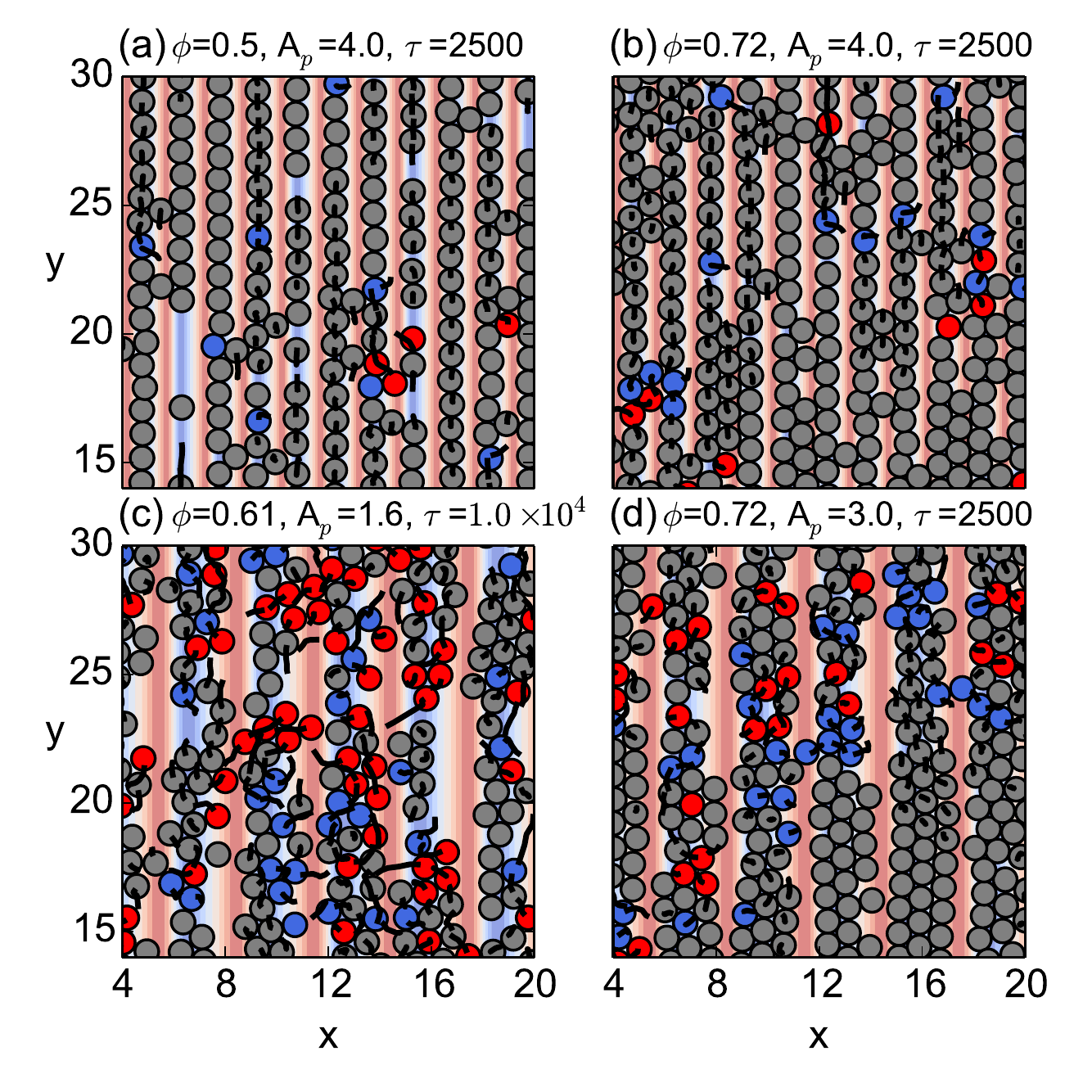}
\caption{ Active particle positions (circles) and trajectories (black lines),
  along with the underlying substrate potential (red and blue lines), in a portion
  of the sample.
  Particles are colored as in Fig.~\ref{fig:3}.
  (a) For $A_p=4.0$,
  $\phi = 0.5$, $a=1.0$, and $\tau = 2500$, the particles form 1D chains
  in each substrate minimum and there
  is a weak normal ratchet effect.
  (b) At $A_p=4.0$, $\phi = 0.72$,
  $a=1.0$, and $\tau=2500$, each substrate minimum contains
  one full row of particles along with a partial second row, and a
  reverse
  ratchet effect occurs.
  (c) Normal ratcheting motion for $A_p=1.6$, $\phi=0.61$,
  $a=2.0$, and $\tau=1\times 10^4$.
  (d) Reverse ratcheting motion for $A_p=3.0$, $\phi=0.72$,
  $a=2.0$, and $\tau=2500$.
}
\label{fig:5}
\end{figure}

In Fig.~\ref{fig:2}(b,e) 
we plot $C_{L}/N$ and $\langle V_{x} \rangle$ versus $\phi$ 
at a higher substrate strength of $A_{p} = 2.0$
for varied $\tau$.
We again find that $C_L/N$ increases with increasing $\phi$ and that
the onset of clustering
drops to lower values of $\phi$ as $\tau$ increases.
For the lowest value of $\tau$, $\langle V_x\rangle=0$, and the maximum
value of
$\langle V_{x} \rangle$ increases with increasing
$\tau$ up to a saturation value of $\tau = 7500$,
above which it decreases with increasing $\tau$, 
indicating that there is a run time that optimizes 
the ratchet efficiency. 
As a function of $\phi$,
$\langle V_{x} \rangle$ is nonmonotonic, starting from a low
value in the single particle limit at low $\phi$,
and then increasing to a maximum value at the optimum
density of $\phi^*=0.55$ before decreasing again.
The value of $\phi^*$ shifts to slightly lower densities as $\tau$ increases.
The nonmonotonic behavior of $\langle V_x\rangle$ arises due to 
a competition between different collective effects.
Since $A_p>F^m$, isolated particles cannot hop over the substrate
barriers in either direction, so that $\langle V_x\rangle=0$ 
at low $\phi$ in the single particle limit.
As $\phi$ increases,  
the increased chance for collisions
permits a collective barrier hopping process to occur
in which interactions between multiple particles permit at least
one of the particles to hop over a substrate barrier, preferentially
in the easy direction of the substrate asymmetry.
For $A_p \leq F^m$,
interactions between pairs of particles is enough to permit motion in the
easy direction to occur, but for $A_p > F^m$,
three-body collisions are required to push at least one particle into a neighboring
substrate minimum, 
preferentially on the easy direction side.  
At low density the three-body interactions cannot occur and
there is no ratcheting effect, 
but as $\phi$ increases 
the magnitude of the normal ratchet effect 
increases up to $\phi \approx 0.5$.  
Above this density, 
the particle-particle interactions overwhelm 
the substrate asymmetry as described above for the $A_p=0.8$ case,
bringing the rate of forward hopping down 
until it equals the rate of backward hopping and
the ratchet effect is lost.

In Fig.~\ref{fig:4}
we show the particles and trajectories for the system in
Fig.~\ref{fig:2}(b,e) with $A_p=2.0$.
At $\phi=0.24$ and $\tau=100$,
shown in Fig.~\ref{fig:4}(a), 
the particles remain confined in the substrate potential minima
and $\langle V_x\rangle=0$. 
At $\phi=0.72$ and $\tau=100$, shown in
Fig.~\ref{fig:4}(b),
we still find $\langle V_{x} \rangle = 0$, 
and the particles form a uniform pinned state.
When clustering first begins at $\phi=0.72$ for $\tau=2500$,
there is a normal ratchet effect illustrated in
Fig.~\ref{fig:4}(c),
while for $\phi=0.72$ and a higher run time of $\tau=1\times 10^5$,
Fig.~\ref{fig:4}(d) indicates that
large numbers of clusters form in the system and interfere with
the normal ratchet effect, reducing its magnitude.

In Fig.~\ref{fig:2}(c,f) 
we show $C_{L}/N$ and $\langle V_{x} \rangle$ versus $\phi$ for a variety of
$\tau$ values
in the strong substrate limit of $A_{p} = 4.0$. 
Here the $C_{L}/N$ curves show little variation with $\tau$ since
the arrangement of the particles
is dominated by the substrate minima.
At the smallest values of $\phi$, the particles are strongly trapped in the
substrate minima and
$\langle V_{x} \rangle = 0$.
For $0.2 <  \phi <  0.6$, we observe a normal ratchet effect with a maximum
efficiency near $\phi = 0.4$.
A reverse ratchet effect appears
for $\phi \geq 0.6$
that is
most pronounced for $\tau = 2500$
and
decreases in magnitude for larger $\tau$.

\begin{figure*}
\includegraphics[width=2\columnwidth]{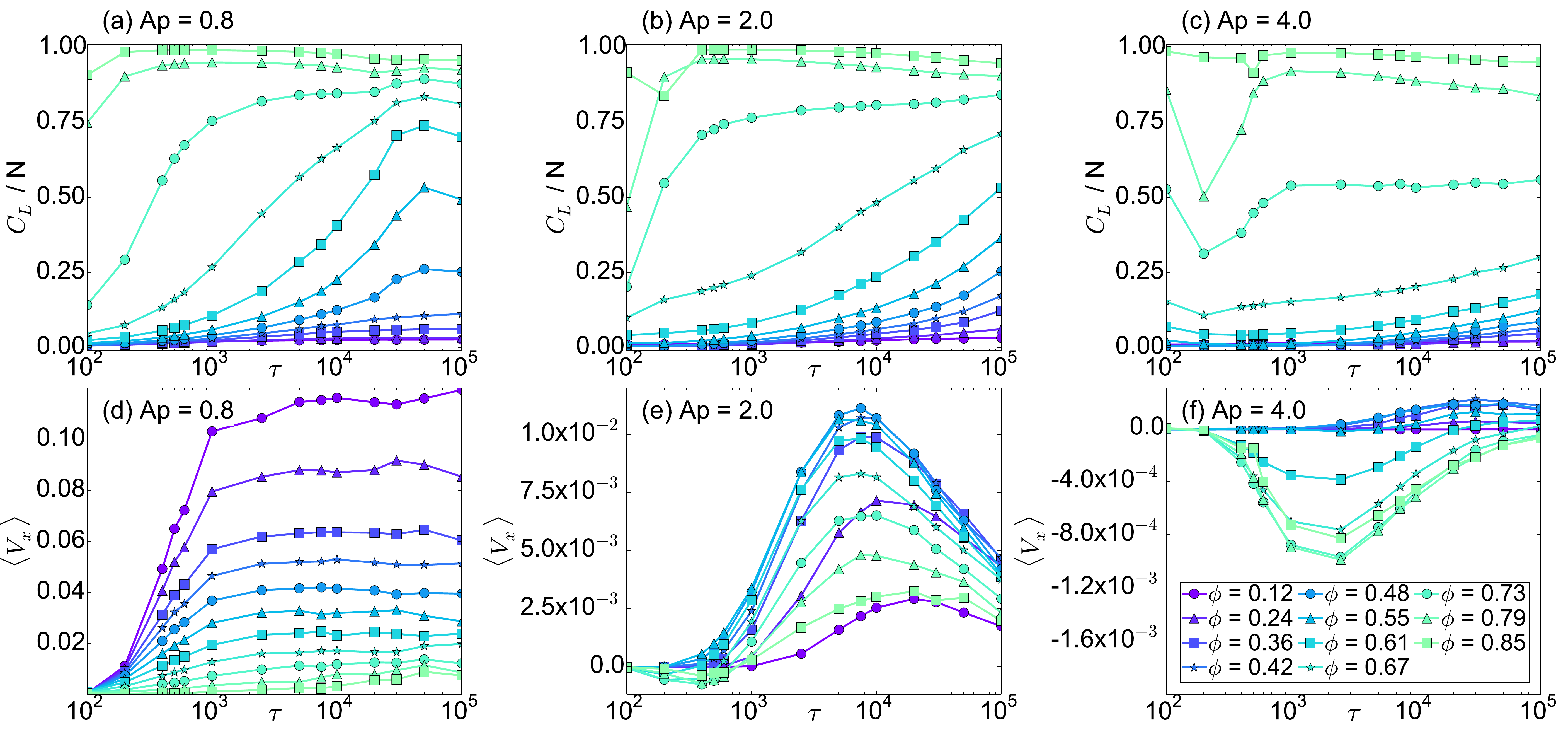}
\caption{(a,b,c) $C_L/N$ vs $\tau$ and (d,e,f) $\langle V_x\rangle$ vs $\tau$
  for $\phi=0.12$, 0.24, 0.36, 0.42, 0.48, 0.55, 0.61, 0.67, 0.73, 0.79, and 0.85.
  (a,d) $A_{p} = 0.8$.
(b,e) $A_{p} = 2.0$.
  (c,f) $A_{p} = 4.0$, where a reverse ratchet effect can occur.
}
\label{fig:6}
\end{figure*}

In Fig.~\ref{fig:5}(a) we show that at
$A_p=4.0$, $\phi = 0.5$, and $\tau = 2500$,
the particles form 1D chains in each substrate minimum
and a weak normal ratchet effect occurs.  
At $\phi=0.72$ and $\tau=2500$,
where a reverse ratchet effect occurs, Fig.~\ref{fig:5}(b) shows that the
particles have buckled out of the bottom of the substrate minima to form
two partial rows of particles.  
The reversal of the ratchet effect
as a function of $\phi$ occurs when the buckling of the particles
in the substrate potential causes particle-particle interactions
instead of particle-substrate interactions to dominate
the particle motion.
When the substrate is strong,
for low $\phi$ each substrate minimum captures a single row of particles.
Although isolated particles show no ratcheting behavior, when pairs of
particles can come into contact along the $y$ direction, one member of the pair
can escape into a neighboring substrate minimum, preferentially on the easy
direction side, producing a weak normal ratchet effect.  For $\phi \geq 0.5$,
a single row of particles can no longer fit in each substrate minima, and the particles
buckle to form one nearly complete row and a second partial row.  The nearly complete row
rests on the hard side of the substrate minimum since this provides a stronger
confinement, while the partial row rests on the easy side of the substrate minimum.
The effective asymmetry of the potential is reversed since the difference in slopes on the
hard and easy sides of the substrate minimum becomes unimportant at these high
densities, and instead, the physical distance to the substrate maximum becomes the
dominant effect.  The maximum is closer to the minimum in the hard direction than in the
easy direction, so hopping in the hard direction is favored at large $\phi$, producing a
reverse ratchet effect.
As $\phi$ is further increased,
the substrate minimum becomes filled with particles and
the asymmetry in the distance to the substrate maximum becomes less important
in determining the hopping direction, causing a decrease in the
efficiency of the reverse ratchet,
as shown in 
Fig.~\ref{fig:2}(f) for $\phi > 0.7$.
Similarly, the reverse ratchet effect becomes very weak at large values of $\tau$
when the local asymmetry in the distance to the substrate maximum becomes
unimportant.

In Fig.~\ref{fig:6} we show 
$C_{L}/N$ and $\langle V_{x} \rangle$ 
as a function of $\tau$ 
for varied $\phi$ at different substrate strengths.
At $A_p=0.8$ in
Fig.~\ref{fig:6}(a),
$C_{L}/N$ increases with increasing $\tau$ 
at low $\phi$ but saturates to $C_L/N \approx 1.0$ at high $\phi$.
In Fig.~\ref{fig:6}(d),
the corresponding $\langle V_{x} \rangle$
curves increase with increasing $\tau$ 
up to a plateau, while the average value of $\langle V_x\rangle$ drops
as $\phi$ increases, indicating that for weak substrates with
$0 < A_p < 1.0$, self-clustering suppresses the normal ratchet effect
by nullifying the substrate asymmetry.
At $A_p=2.0$ in Fig.~\ref{fig:6}(e),
$\langle V_{x} \rangle$ is nonmonotonic, with a peak in the
magnitude of the ratchet effect
for $5000 < \tau < 1\times 10^4$, and a decrease in
$\langle V_{x} \rangle$ at large values of $\tau$ due to the
occurrence of self-clustering.
For high densities of $\phi>0.67$, there
is a window at small $\tau$ in which a reverse ratchet effect appears
caused by particles hopping over the closest maximum instead of traveling
up the least steep
side of the potential.
Figure~\ref{fig:6}(c,f) shows
$C_{L}/N$ and $\langle V_{x} \rangle$ versus $\tau$ 
for 
$A_{p} = 4.0$. 
There is a normal ratchet effect at large $\tau$ for
$\phi < 0.55$ 
and a reverse ratchet effect at intermediate $\tau$ for
$\phi \geq 0.55$.
The efficiency of the reverse ratchet is highest at $\tau \approx 5000$.

\begin{figure*}
\includegraphics[width=2\columnwidth]{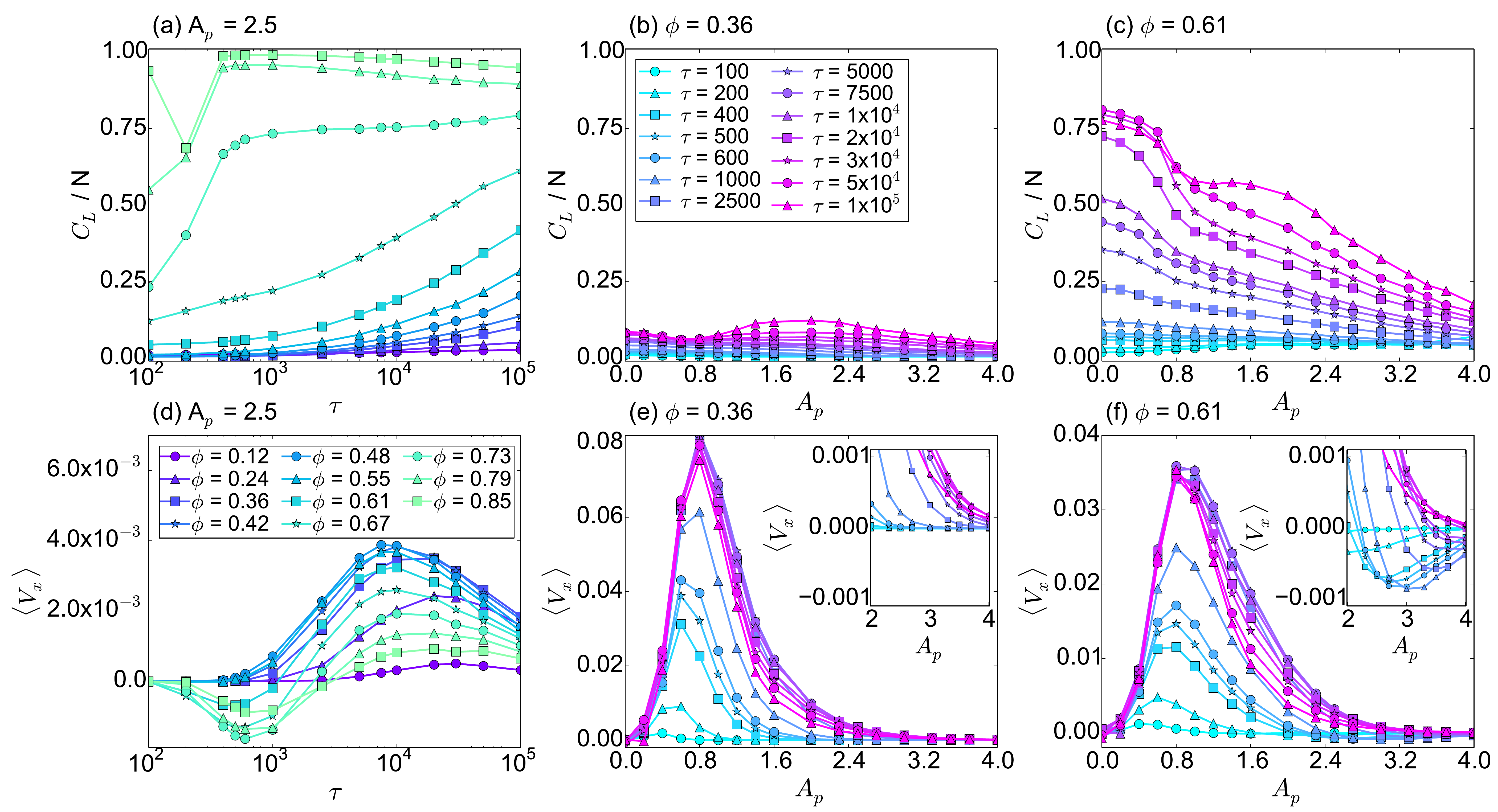}
\caption{(a) $C_{L}/N$ vs $\tau$ and
  (d) $\langle V_{x} \rangle$ vs $\tau$
  for $\phi=0.12$, 0.24, 0.36, 0.42, 0.48, 0.55, 0.61, 0.67, 0.73, 0.79, and 0.85
  at $A_p=2.5$, where there can be a transition from a reverse ratchet to a
  normal ratchet with increasing $\tau$.
  (b,e)
  $C_{L}/N$ and $\langle V_{x} \rangle$ vs $A_{p}$ at $\phi = 0.36$ for
  $\tau=100$, 200, 400, 500, 600, 1000, 2500, 5000, 7500, $1\times 10^4$,
  $2\times 10^4$, $3\times 10^4$,
  $5\times 10^4$, and $1\times 10^5$, showing that only a normal ratchet effect occurs.
  (c,f) $C_{L}/N$ and $\langle V_{x} \rangle$ vs $A_{p}$ at $\phi = 0.61$ for
  the same $\tau$ values as in panels (b) and (e).
  Inset of (f): blow up of main panel showing a detail of the weak reverse ratchet
effect that appears at large $A_p$.}
\label{fig:7}
\end{figure*}

For substrate strengths of $1.6 < A_{p} < 4.0$, 
a crossover from a reverse to a normal ratchet effect can occur
as a function of $\tau$, as highlighted in
Fig.~\ref{fig:7}(a,d), 
where 
we plot $C_{L}/N$ and $\langle V_{x} \rangle$ versus $\tau$
for a sample with
$A_p=2.5$.
There is a clear 
transition from a reverse ratchet
to a normal ratchet effect with increasing $\tau$, 
suggesting that in a system composed of 
two species of particles with different $\tau$, 
it could be possible
to have the two species exhibit a net drift
in opposite directions. 

\begin{figure*}
\includegraphics[width=2.0\columnwidth]{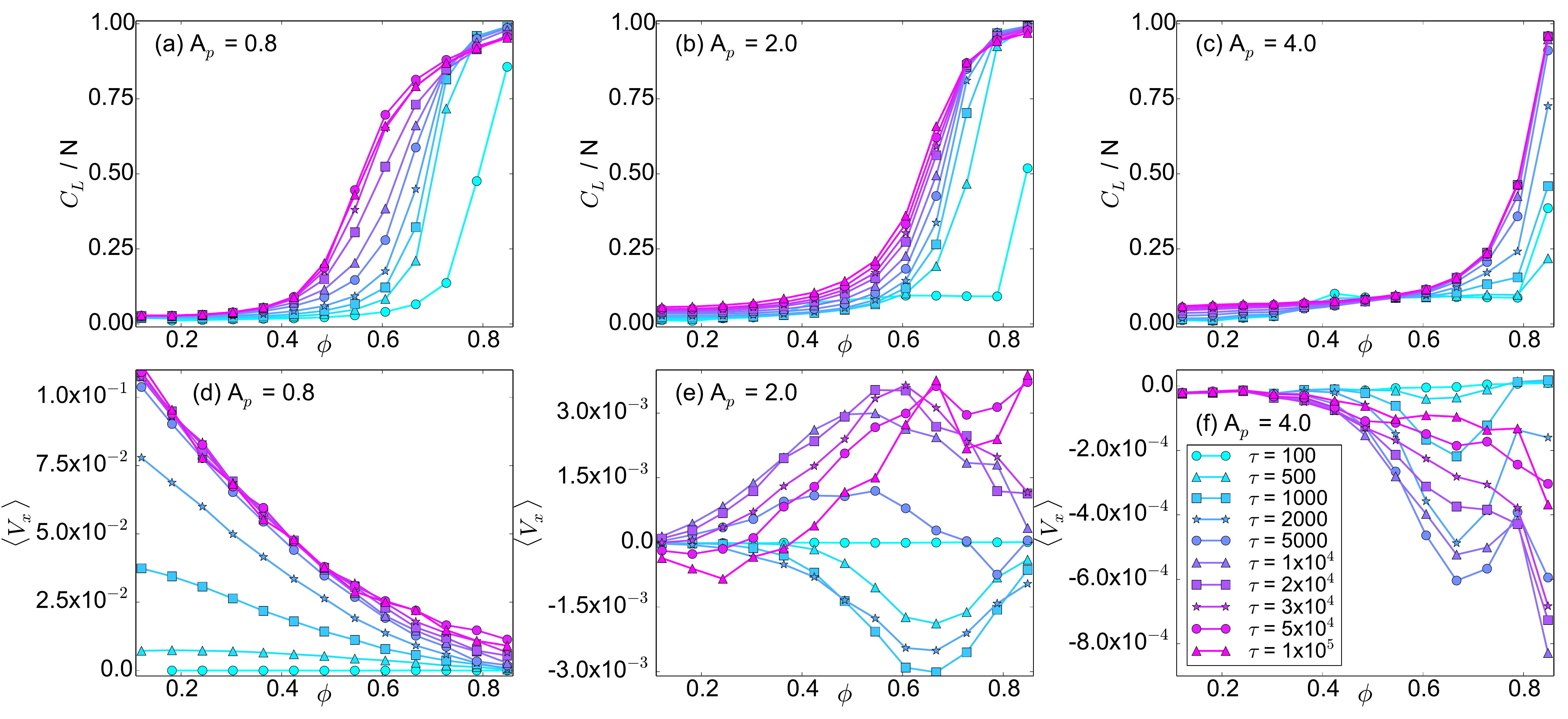}
\caption{(a,b,c) $C_{L}/N$ vs $\tau$ and (d,e,f) $\langle V_{x} \rangle$ vs $\phi$
  for $\tau=100$, 500, 1000, 2000, 5000, $1\times 10^4$,
  $2\times 10^4$, $3\times 10^4$, $5\times 10^4$, and $1\times 10^5$
  in samples with a substrate lattice constant of $a=2.0$, twice as large as the lattice
  constant considered previously.
  (a,d) $A_{p} = 0.8$.
  (b,e) At $A_p=2.0$ a reverse ratchet effect occurs.
  (c,f) At $A_p=4.0$, the magnitude of the reverse ratchet effect
  initially increases with increasing $\tau$ for $\tau<1\times 10^4$ and then
  decreases as $\tau$ further increases.
}
\label{fig:8}
\end{figure*}

\begin{figure*}
\includegraphics[width=2\columnwidth]{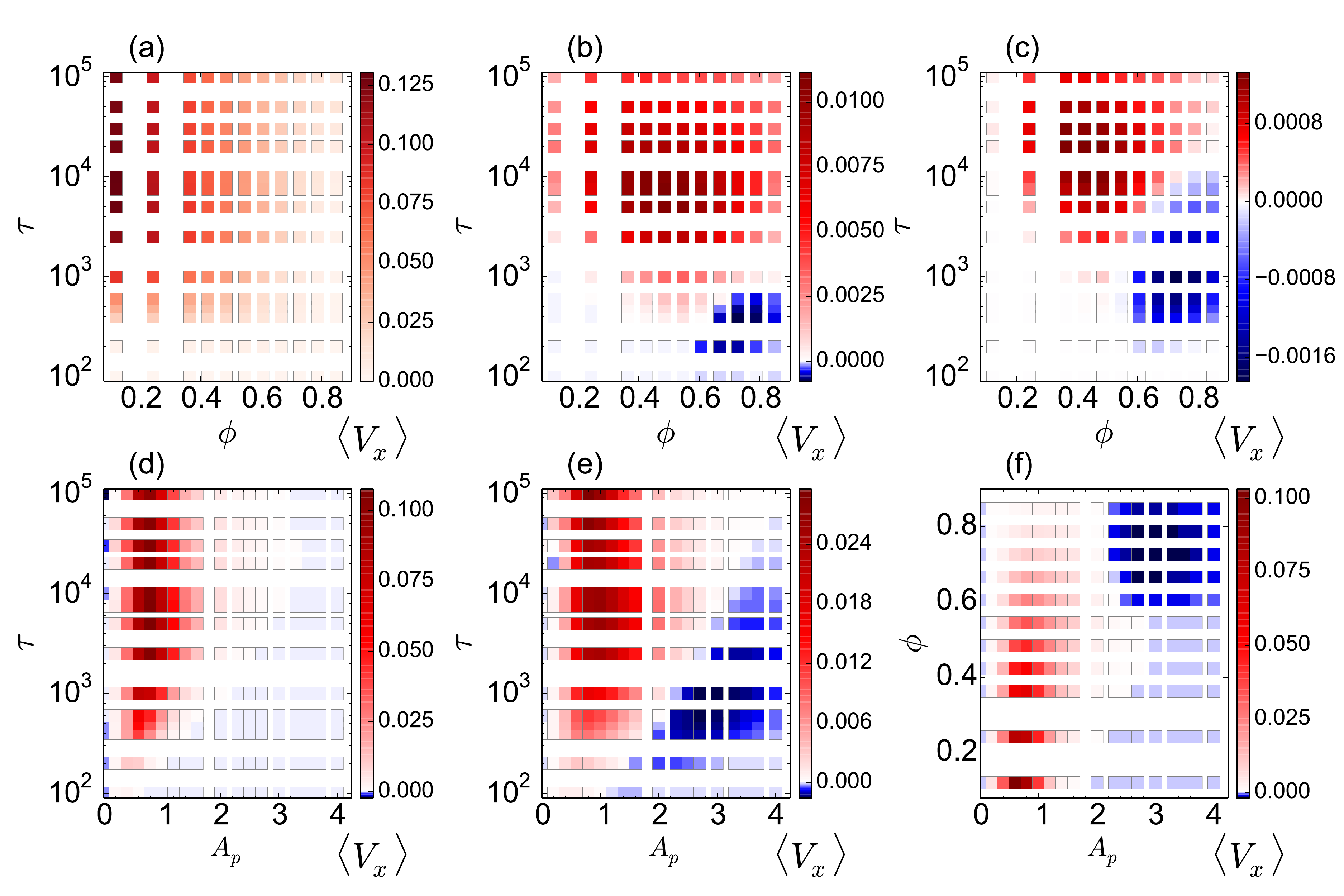}
\caption{ Phase diagrams
  in samples with $a=1.0$ showing the magnitude
  of the ratchet effect
  as determined by the value of $\langle V_x\rangle$,
  with blue denoting a reverse ratchet and red denoting a
  normal ratchet, as indicated by the color bar keys.
  (a,b,c) $\tau$ vs $\phi$ phase diagrams at
  (a) $A_{p} = 0.8$, (b) $A_{p} = 2.0$, and (c) $A_{p} = 3.0$.
  (d,e) $\tau$ vs $A_p$ phase diagrams at
  (d) $\phi = 0.24$ and 
  (e) $\phi = 0.68$.
  (f) $\phi$ vs $A_{p}$ phase diagram at $\tau = 1000.$
}
\label{fig:9}
\end{figure*}

In Fig.~\ref{fig:7}(b,e) we show
$C_{L}/N$ and $\langle V_{x} \rangle$ versus $A_{p}$ 
for varied $\tau$ at 
$\phi = 0.36$,
while in Fig.~\ref{fig:7}(c,f) we show the same measures
at $\phi=0.61$.
In Fig.~\ref{fig:7}(b) at $\phi=0.36$, the particle density is
too low for large clusters to appear,
while in Fig.~\ref{fig:7}(c) at $\phi=0.61$,
the clustering that occurs for low $A_p$ is suppressed as the
substrate strength is increased.
At $A_p=0$ in Fig.~\ref{fig:7}(e,f), 
$\langle V_{x} \rangle=0$,
and as $A_p$ increases, $\langle V_x\rangle$ increases to a
maximum value near $A_p=0.75$ before decreasing back to zero at
higher $A_p$.
The normal ratchet effect operates most efficiently when particles can overcome
only the barrier for motion in the easy direction, and not the barrier in the hard direction.
The particles can overcome the barrier in the easy direction when
$A_p \leq A_p^e$ with $A_p^e=(3/2)F^m$, and they
can overcome the barrier in the hard direction when
$A_p \leq A_p^h$ with $A_p^h=(3/4)F^m$.
Thus, the ratchet effect is zero for large $A_p$, becomes finite
below $A_p = A_p^e =1.5$, and diminishes rapidly
below $A_p \approx A_p^h=0.75$.
In Fig.~\ref{fig:7}(c,f) at $\phi = 0.61$, increasing $A_{p}$
decreases the cluster size $C_L/N$
for $\tau<7.5\times 10^4$ as the particles become increasingly localized,
while at $\tau=1\times 10^5$, 
$C_L/N$
develops a small nonmonotonic peak near $A_p=1.5$.
For all $\tau$, 
there is still a local maximum in $\langle V_{x} \rangle$ near $A_{p} \approx 0.75$, while
for $A_p>2.0$ at low values of $\tau$, a weak
reverse ratchet effect occurs

We have also examined systems with different substrate periods $a$.
We find that for larger values of $a$, 
the onset of
the formation of multiple rows of particles in a single substrate minimum shifts to
lower $\phi$, and that therefore
the reverse ratchet regime is enhanced.
In Fig.~\ref{fig:8}(a,d) we show 
$C_{L}/N$ and $\langle V_{x} \rangle$ versus $\phi$ 
at $A_{p}= 0.8$ for varied $\tau$ in a sample with $a=2.0$, twice as
large as the lattice constant considered previously.
For $A_{p} < 1.0$, 
there is a normal ratchet effect that generally decreases in magnitude with
increasing $\phi$ and that saturates in magnitude with increasing $\tau$.
In Fig.~\ref{fig:8}(b,e), we plot the same quantities for $A_{p} = 2.0$,
where the ratchet effect becomes nonmonotonic and switches from a reverse
ratchet 
for $\tau < 5000$ to a normal ratchet  for $\tau \geq 5000$.
In contrast, 
for the $a=1.0$ system in Fig.~\ref{fig:3} 
at the same value of $A_{p}$,  
there is almost no reverse ratchet regime.
In Fig.~\ref{fig:8}(c,f),
the $C_L/N$ and $\langle V_x\rangle$ versus $\phi$
curves for $a=2.0$ at $A_{p} = 4.0$ indicate that  
the ratchet effect is always in the reverse direction with a magnitude that
increases
with increasing $\phi$, a behavior that is the opposite of that observed
at $A_{p} = 0.8$.

\begin{figure}[h!]
\includegraphics[width=\columnwidth]{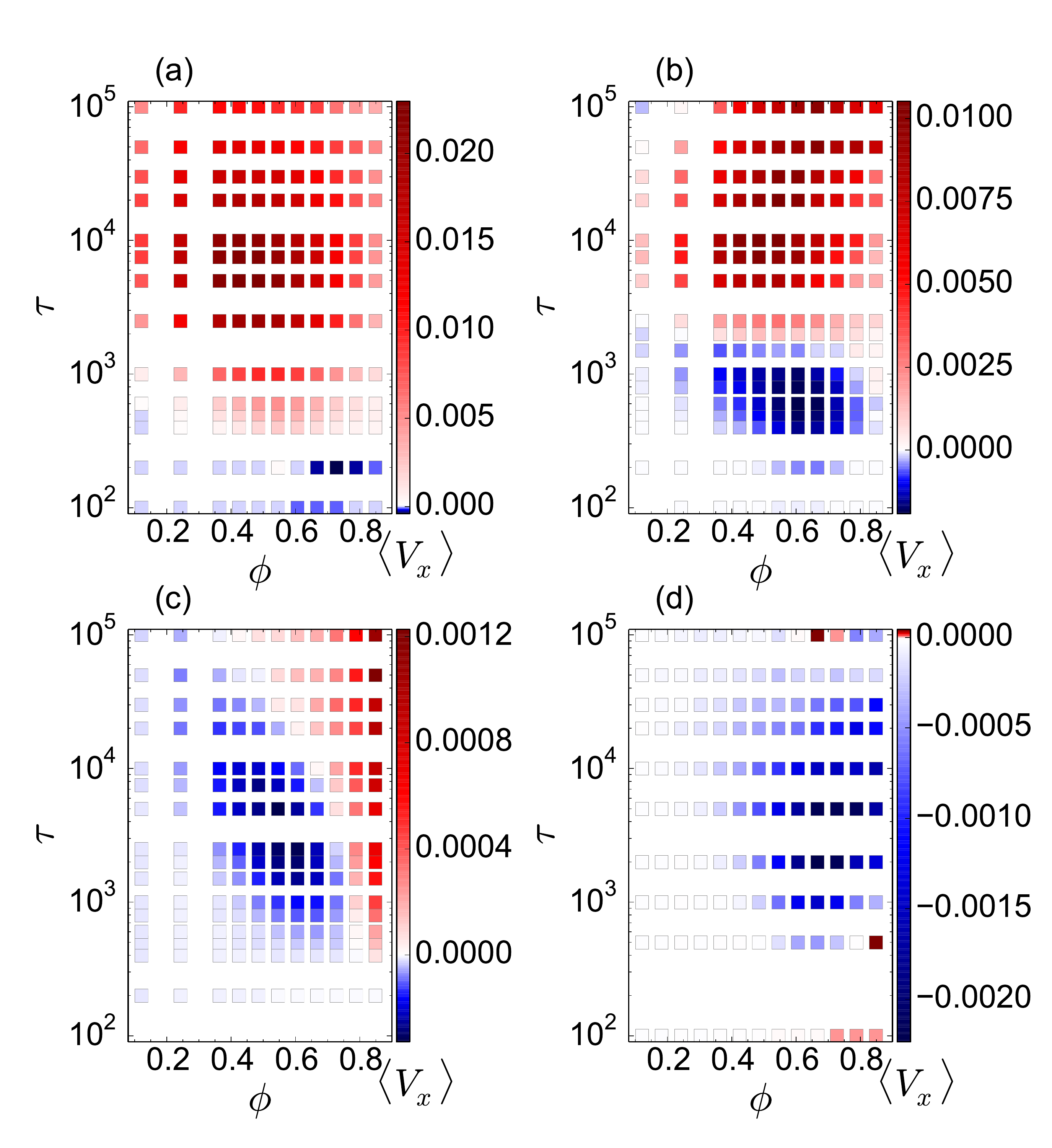}
\caption{ Phase diagrams
  as a function of $\tau$ vs $\phi$
  showing the magnitude of the ratchet effect as determined by
  the value of $\langle V_x\rangle$, with blue denoting a reverse ratchet and red
  denoting a normal ratchet, as indicated by the color bar keys.
  (a)
  At $A_{p} = 1.6$ and $a = 1.0$,
there is a normal ratchet effect.
(b) At $A_p=1.6$ and $a=2.0$,
there is a transition from a normal to a reverse ratchet effect.
(c)
At $A_{p} = 3.0$ and $a = 1.5$,
the reverse ratchet regime extends to
higher values of $\tau$.
(d)
At $A_p=3.0$ and $a = 2.0$,
the ratchet effect is predominately in the reverse direction.
}
\label{fig:10}
\end{figure}

In Fig.~\ref{fig:9}
we highlight the different ratcheting behaviors for $a=1.0$ in a series of phase diagrams 
colored according to the value of $\langle V_{x} \rangle$, 
where blue indicates a reverse ratchet effect,
white indicates
no ratchet effect,
and red indicates
a normal ratchet effect.
In Fig.~\ref{fig:9}(a) we show a
$\tau$ versus $\phi$ phase diagram at $A_{p} = 0.8$ where only a normal
ratchet effect occurs.
The ratchet efficiency
decreases with decreasing $\tau$ and increasing $\phi$.
At $A_p=2.0$ in Fig.~\ref{fig:9}(b),
the $\tau$ versus $\phi$ phase diagram plot indicates that
the maximum
normal ratchet effect appears
near $\phi = 0.5$ and $\tau = 10^4$, 
while reverse ratchet effects appear
in the lower right hand corner at low $\tau$ and large $\phi$.
In Fig.~\ref{fig:9}(c), the $\tau$ versus $\phi$ phase diagram shows that
the extent of the reversed ratchet regime is larger and that there is
a clear transition from a reverse to a normal ratchet as a function of
increasing $\tau$ and/or decreasing $\phi$.
As $A_{p}$ is further increased, 
the reversed ratchet region grows,
but the magnitude of the ratchet effect
is generally reduced.
In the $\tau$ versus $A_{p}$ phase diagram at $\phi = 0.24$
in
Fig.~\ref{fig:9}(d),
the ratchet effect is always in the normal direction and
is maximum in a band along the
$A_{p} = 0.8$ line. 
Regions of normal and reverse ratcheting appear in the
$\tau$ versus $A_p$ phase diagram for $\phi=0.68$, as shown in
Fig.~\ref{fig:9}(e).
In the  $\phi$ versus $A_{p}$ phase diagram for fixed $\tau = 1000$ in
Fig.~\ref{fig:9}(f),
a normal ratchet effect occurs for
$A_{p} < 2.0$,  while there is a transition to a reverse
ratchet effect for $A_{p} > 2.0$ and $\phi > 0.5$.

In Fig.~\ref{fig:10}(a) we show the ratchet phase diagram as a function
of $\tau$ versus $\phi$ for a system with $A_{p} = 1.6$ and $a = 1.0$,
where the ratchet effect is always in the normal direction.
For $A_p=1.6$ and $a=2.0$ in Fig.~\ref{fig:10}(b),
at large $\tau$ a normal ratchet effect occurs, as shown in
Fig.~\ref{fig:5}(c) for $\phi=0.61$ and $\tau=1 \times 10^4$, while at
lower $\tau$,
multiple rows of particles can be trapped in
each substrate minimum, producing
a region of reverse ratchet effect.
At higher values of $A_{p}$,
the size of the reverse ratchet effect region increases,
as shown in Fig.~\ref{fig:10}(c) 
for
$A_{p} = 3.0$ and $a = 1.5$.
In Fig.~\ref{fig:10}(d) at $A_p=3.0$ and $a=2.0$,
the ratchet effect is predominately in the reverse direction,
as illustrated in Fig.~\ref{fig:5}(d) for $\phi=0.72$ and $\tau=2500$.

\vspace{0.1in}

\section{Summary}

We show that in a 2D system of 
sterically interacting run-and-tumble disk-shaped particles in the presence
of a quasi-one-dimensional asymmetric periodic substrate, 
a variety of collective active ratchet behaviors can occur, 
including nonmonotonic
changes in the ratchet effect magnitude as well as ratchet reversals.
A normal ratchet effect, where a net drift of the particles occurs
along the easy flow direction of the substrate,
appears for weak substrates,
and the ratchet efficiency generally decreases 
with increasing particle density due to self-jamming
or clustering effects since it is more difficult for a cluster to
undergo ratcheting motion than for individual particles
to do so.
For intermediate substrate strengths,
the ratchet efficiency is nonmonotonic
as function of particle density or activity. 
At low particle densities, 
individual particles cannot jump over the potential barrier, 
but particle-particle interactions can produce a
collective particle hopping 
in the easy direction of the substrate asymmetry. 
When the particle density or activity is high enough, 
strong clustering effects occur that reduce the ratchet efficiency.
For strong substrates where 
multiple rows of particles can form in each substrate minimum,
it is possible to realize a reverse ratchet effect in which
the net flux of particles is in the hard direction of the substrate asymmetry.
The size of the reverse ratchet regime can be increased by increasing the
size of the substrate periodicity,
which shifts the transition from single to multiple rows of particles per
substrate minimum to lower particle densities.
Our work shows that by exploiting collective effects, 
it is possible to create reversible active matter ratchets which 
could be useful  
for various sorting applications.

\section{Acknowledgements}
This work was carried out under the auspices of the 
NNSA of the 
U.S. DoE
at 
LANL
under Contract No.
DE-AC52-06NA25396.
The work of DM was supported in part by the U.S. DoE, Office of Science,
Office of Workforce Development for Teachers and Scientists (WDTS) under the
Visiting Faculty Program (VFP).

\footnotesize{
 
}
\end{document}